# Light Emission by Free Electrons in Photonic Time-Crystals

**Alex Dikopoltsev[1][†], Yonatan Sharabi[1][†], Mark Lyubarov[1], Yaakov Lumer[1], Shai Tsesses[2],**

**Eran Lustig[1], Ido Kaminer[2], and Mordechai Segev[1,2]**

[1]Physics Department, Technion – Israel Institute of Technology, Haifa 32000, Israel
[2]Electrical Engineering Department, Technion – Israel Institute of Technology, Haifa 32000, Israel

[†]*These authors contributed equally*

**Photonic time-crystals (PTCs) are spatially-homogeneous media whose electromagnetic (EM) susceptibility varies periodically in time, causing temporal reflections and refractions for any wave propagating within the medium. The time-reflected and time-refracted waves interfere, giving rise to Floquet modes with momentum bands separated by gaps (rather than energy bands and gaps, as in photonic crystals). Here, we show that a free electron moving in a PTC spontaneously emits radiation, and, when associated with momentum-gap modes, the radiation of the electron is exponentially amplified by the modulation of the refractive index. Moreover, under strong electron-photon coupling, the quantum formulation reveals that the spontaneous emission into the PTC bandgap experiences destructive quantum interference with the emission of the electron into the PTC band modes, leading to suppression of the interdependent emission. Free-electron physics in PTCs offers a new platform for studying a plethora of exciting phenomena such as radiating dipoles moving at relativistic speeds and highly efficient quantum interactions with free electrons. Radiation emission in PTCs and its non-resonant nature hold the promise for constructing new particle detectors with adjustable sensitivity and highly tunable laser sources, ranging from terahertz to the x-ray regime, drawing their power from the modulation.**

Photonic time-crystals (PTCs) are materials with a time-periodic change of their electromagnetic (EM) properties. A sudden change in the optical properties leads to a time-reflection and time-refraction for any EM wave traveling within that medium [1–3]. Since causality dictates that time cannot be reversed, the reflected wave generated by the abrupt temporal variation (henceforth the time-reflected wave) cannot go back in time; instead, time-reflections go back in space, with a conjugate phase. Making a periodic sequence of changes in the material properties causes the forward-propagating (time-refracted) waves and the time-reflections to interfere, leading to the emergence of a band structure and dispersion relation [4,5]. Naturally, PTCs seem similar to one-dimensional photonic crystals (PCs): the dispersion relations of the PTC and the photonic crystal are both determined by the strength and periodicity of the change in the EM properties. However, there are several profound fundamental differences between the two. First, while dielectric spatial photonic crystals conserve energy and do not conserve momentum, in PTCs the temporal variations break time-translation symmetry, and therefore energy is not a conserved quantity in PTCs. On the other hand, being spatially homogenous, PTCs conserve momentum. The conservation of momentum means that the $k$-wavevector is a good quantum number in the PTC [6]. Second, the roles of $\omega$ and $k$ are swapped in the dispersion curve of a PTC, meaning the bandgaps in the PTC are in momentum rather than in energy [6,7]. The third difference has to do with the modes associated with the bandgap. A wave arriving at an air/dielectric-PC interface with a frequency associated with the PC bandgap experiences total reflection, because the PC has no propagating modes that can support such a frequency. In PTCs, on the other hand, because energy is not conserved, waves with wavevectors inside the momentum-gap change the energy they carry, and exhibit not only exponential decay but also exponential growth in time [8,9].

To observe the main features of a PTC, such as the momentum-gap [8], the modulation amplitude of the EM properties must be large (of order unity) and the modulation frequency should be high enough – on the order of the frequency of the EM wave propagating within the PTC. Thus far, PTCs were demonstrated only at radio frequencies in electronic transmission lines [7]. However, the transition to optical frequencies is near, as PTCs are now attracting growing attention due to recent advances in fabricating dynamic optical systems and metamaterials. Recent experiments in epsilon-near-zero materials [10] showed large (on the order of ~1) femtosecond scale variations in the refractive index [11,12]. These and later experiments of time-refraction [13,14] herald the possibility to experiment with photonic time crystals in the near future [15,16].

Here, we study the interaction of free electrons with PTCs. We show, through classical and quantum electrodynamic formulations, that free electrons propagating in a PTC spontaneously radiate due to the exchange of energy with their time-varying surrounding. The emission exhibits a tunable spectrum in two different regimes, subluminal and superluminal, displaying an abrupt transition in the shape of the radiation pattern when the electron becomes faster than the phase velocity in the medium and crosses the Cherenkov radiation threshold. The time-varying medium contributes an energy "kick" to the interaction, similar to Compton scattering, yet with the medium acting as the scattering photon. For EM waves with wavevectors $k$ residing in the momentum-gap, this energy "kick" leads to exponential amplification of the radiation, drawing the energy from the modulation. The quantum electrodynamic analysis of the system with strong electron-photon coupling reveals a new effect: the quantum degeneracy between the final states of the emission process by the moving electron and the states of the spontaneous emission created by vacuum fluctuations in the PTC is lifted by the interaction term, and causes avoided-crossing between the two processes. This causes the spontaneous emission into the momentum bandgap to display destructive quantum interference with the emission of the electron into PTC band modes, giving rise to suppression of the emission at the crossing point. The process of free-electron emission in PTCs, and especially the exponential enhancement of the emission driven by the temporally-modulated refractive index, offer a plethora of new phenomena and suggest novel applications such as widely tunable lasers ranging from the terahertz regime all the way to X-rays, drawing their energy from the modulation.

Henceforth, we consider the radiation of an electron traveling in a PTC medium with $\epsilon(t) = \epsilon(t + T)$. The general geometry is presented in Fig 1a, showing an electron moving at a constant velocity inside a spatially-homogeneous PTC while emitting radiation. Figure 1b presents a general outcome of an FDTD simulation of Maxwell's equations, displaying the electron radiation in a PTC of sinusoidal modulation $\epsilon(t) = \epsilon_0 + \epsilon_1 \sin(\Omega t)$ starting at $t = T_1$ and ending at $t = T_2$ (the time-analogue of passing through a 1D photonic crystal of finite length). Importantly, this simulation is carried out for the case where the electron velocity is below the Cherenkov threshold, namely, the electron moves slower than the speed of light in the medium. Nevertheless, as Fig. 1b highlights, the temporal variations responsible for the PTC enable free electron radiation even in the regime that Cherenkov radiation cannot exist [17,18]. Moreover, since the PTC medium is homogeneous, all effects associated with a spatial periodicity (the Smith-Purcell effect [19]) do not exist either. Yet the free electron displays efficient radiation, which is actually angle-dependent despite the fact that the medium is homogenous. Namely, as shown in Fig 1b, waves at higher wavenumbers are emitted forward, closer to the propagation axis, while waves at lower wavenumbers are emitted backwards. The results of this direct simulation immediately indicate that a free electron passing through a PTC radiates in a spatio-temporal pattern that is fundamentally different from known Cherenkov radiation.

To find the radiation of moving electrons in a PTC, we first analyze the EM modes in this time-modulated medium using Maxwell's equations. Analogously to an abrupt change in space, a temporally-abrupt change of the permittivity leads to time-reflection and time-refraction, both shifted in frequency [4]. However, time-reflections cannot occur backwards in time, because (to the best of our knowledge) time only moves forward. Hence, time-reflections, even when generated by a spatially-uniform temporal change in the medium, actually occur in space and appear as spatial back-reflections. Naturally, when the susceptibility in a medium undergoes periodic changes in time, the refractive index variations induce multiple reflections and refractions of waves in space. Due to the periodicity of the variations, the reflections and refractions interfere to form a Floquet band structure of the EM modes (Fig. 2a). When the medium is homogeneous, the wavevector **k** is a conserved quantity appearing as a constant characterizing each eigenmode. The Floquet modes of the EM field are best expressed through a magnetic field [20] of the form $\mathbf{H_k}(\mathbf{r},t) = \mathbf{H}_0 e^{i\mathbf{k}\mathbf{r}} h_k(t); \; h_k(t) = e^{i\omega_k t}\sum q_\mathbf{k}^m e^{im\Omega t}$, where $\omega_k$ is the Floquet frequency that depends on the wavenumber $k$, $\Omega$ is the modulation frequency $2\pi/T$ ($T$ is the period), $m$ is the order of the harmonic, and $q_\mathbf{k}^m$ are the coefficients of the Floquet mode harmonic. Notice that, for each wavenumber $k$ there are two Floquet function solutions, $h_{k,+}(t)$ and $h_{k,-}(t)$, for $\omega_k$ and $-\omega_k$, respectively. When $k$ is in the band, these functions are complex conjugates, $h_{k,+}(t)=h_{k,-}^*(t)$. But, when $k$ is in the gap $\omega_k$ becomes complex, therefore, one of the solutions is decaying exponentially and one is growing exponentially with time (see SI). For these momentum-gap modes, the energy carried by the modes grows or decays exponentially, at the expense of the energy invested in the temporal modulation of the permittivity. The superposition of all modes in the bands and in the gaps, each with its own wavenumber **k**, describes the EM surrounding of the electron passing through the PTC.

We model the electron classically, as a point-charge moving in the $z$-direction, with velocity $v = \beta c_0$, creating electric current density $\boldsymbol{j}_e(\boldsymbol{r},t) = \delta(r_\perp)\delta(z - \beta c_0 t)\hat{z}$, where $c_0$ is the speed of light in vacuum, $0 < \beta < 1$ is a real number, and $r_\perp$ are the coordinates transverse to $z$. The point source can be decomposed into its spatial wavenumbers $\boldsymbol{j}_k(\boldsymbol{r},t) = j_0 e^{ik_z \beta c_0 t}\hat{z}$. The wave equation for the magnetic field component $\boldsymbol{H}_k$ in a PTC with a current source $\boldsymbol{j}_k$ is

$$(\partial_t(\epsilon_m(t)\partial_t) - c^2 k^2)\boldsymbol{H}_k(t) = -i\boldsymbol{k} \times \boldsymbol{j}_k \qquad (1),$$

where $c = 1/\sqrt{\mu_0 \epsilon_0 \epsilon_{r,0}}$, and the permittivity is $\epsilon(t) = \epsilon_0 \epsilon_{r,0} \epsilon_m(t)$ where $\epsilon_m(t)$ is the time dependent part and $\epsilon_m(0) = 1$. To solve this equation, we find the Green function which satisfies the time-dependent part of Eq. 1, $(\partial_t(\epsilon_m(t)\partial_t) - c^2 k^2)G_k(t,t') = \delta(t - t')$. Using the time dependent part of the magnetic field, $h_k(t)$, we find (details in the SI)

$$G_k(t,t') = \begin{cases} \frac{h_{k,+}(t')h_{k,-}(t) - h_{k,-}(t')h_{k,+}(t)}{C_k} & t < t' \\ 0 & t \geq t' \end{cases} \qquad (2),$$

where $C_k = \epsilon(t)\left(h_{k,+}(t)\dot{h}_{k,-}(t) - h_{k,-}(t)\dot{h}_{k,+}(t)\right)$ is constant in time, which, interestingly, is proportional to the Minkowski momentum, $\int (\boldsymbol{D} \times \boldsymbol{B})dV$ [21]. The Minkowski momentum is a conserved quantity for a homogenous medium even when the medium varies in time, and is used here to define the amplitude of each mode. By integrating over the product of the Green function and our current source, $\boldsymbol{j}_k$, we find the EM radiation induced by the electron. The radiation efficiency of the electron is determined by the relation between its phase $k_z \beta c_0 t$ and the phase of the Floquet modes $h_k(t)$, a necessary requirement for efficient free-electron interaction with matter. Maximum efficiency occurs when the phases match, that is, when $k_z \beta c = \omega_k + m\Omega$, which occurs for every wavenumber **k** and harmonic number $m$ separately. These considerations allow for phase matching of the radiation process in time and the z-direction simultaneously, which results in the following magnetic field component

$$H_{k,y/x,m}(t) = -i\frac{k_{x/y}h_k(t)q_k^{m*}t}{2C_k} \quad (3).$$

We find that the amplitude of the field grows with time and is proportional to the strength of the relevant Floquet harmonic, $|a_k^m|$. Moreover, when the wavenumber of the source resides in the momentum-gap, any coupling to that gap mode induces exponential growth of the radiation amplitude, in line with the exponential growth of $h_{k,+}(t)$ (See the analytical derivation in the SI).

To find the exact angles and frequencies to which the electron radiates, we map the phase-matching condition onto the wavevector space $(k_z, k_\perp)$ of radiation in the PTC

$$k_\perp^2 = k_z^2\left(n_{eff}^2(k)\beta^2 - 1\right) - 2mk_z\beta n_{eff}k_{mod} + m^2k_{mod}^2, \quad (4),$$

where $n_{eff}(k) = \omega_k/kc_0$ and $k_{mod} = \Omega n_{eff}/c_0$. This mapping is natural because the wavevector **k** is a conserved quantity for each mode in our system. Figures 2b and 2c show this mapping in two distinct cases, where the electron is below and above the Cherenkov velocity threshold, respectively. The Cherenkov threshold ($\beta n_r = 1$) plays a significant role in this interaction – it differentiates between two regimes of radiation, subluminal and superluminal. We notice that the main harmonic ($k_z\beta c = \omega_k$; m=0) is missing in the subluminal regime, where $\beta = 0.4 < 1/n_r$ (Fig. 2b). This is because the electron is not fast enough on its own to radiate, for the very same reason there is no Cherenkov radiation below the threshold. However, because the medium here is a PTC, the temporal modulation of the permittivity endows the electron with energy to interact with frequencies of integer quanta $\Omega$ higher and lower than the Floquet frequency $\omega_k$ of the radiated modes. For example, when phase-matching is met for harmonic $m = -1$, with $k_z\beta c =$

$\omega_k - \Omega$, radiation is efficiently emitted into this Floquet mode with its original frequency $\omega_k$. On the other hand, in the superluminal regime (Fig. 2c) with $\beta = 0.99 > 1/n_r$, the electron is fast enough not only to emit ordinary Cherenkov radiation, which is here modified by the PTC dispersion ($m = 0$), but also to radiate to higher harmonics with $m > 0$. Morever, in the superluminal regime, the phase-matching condition is fulfilled also for gap modes. The coupling varies according to the shape of the bandgap modes (see SI), but after being excited by the electron - all gap radiation grows exponentially.

To confirm our analytical results, we simulate this system with an FDTD algorithm. Figure 2d shows a comparison between the magnetic field amplitude found in simulations and the analytic calculation, for an electron moving with $\beta = 0.4$ in a medium with permittivity $\epsilon(t) = 2 + 0.2 \sin(\Omega t)$, at the end of of 50 PTC cycles. The simulation results conform well for the full range of angles (up to the resolution and system size limits of FDTD simulations). Figures 2e and 2f display the Fourier components of the output magnetic field for $\beta = 0.4$ and $0.99$, respectively (same as in Fig. 2b and 2c), when $\epsilon_1/\epsilon_0 = 0.2$. These simulations show that the electron radiation indeed follows the phase-matching condition presented above for the Floquet band modes. But even more interestingly, when the electron radiates into phase-matched modes within the bandgap, the radiation grows exponentially in time (analytic expression given in the SI [22]).

Thus far, we treated the free electron as a classical moving point charge, and showed that the emitted EM radiation strongly depends on the band structure of the PTC. In the subluminal (spectrally limited) case ($\beta n < 1$), the PTC serves as an "enabler" for the electron radiation, where the free-electron is phase-matched to harmonics with $m < 0$. In the superluminal regime ($\beta n > 1$), the free-electron is phase-matched with all the harmonics of the PTC modes, so in addition to the fundamental harmonic ($m = 0$), which corresponds partially to the ordinary Cherenkov radiation in a stationary medium, we find superluminal radiation in a slew of frequencies and angles. The classical model we use predicts these outcomes, yet, it is incomplete in explaining the free-electron emission to momentum-gap modes This is because – as we show below - a medium that is modulated in time also displays spontaneous photon-pair creation exactly in the momentum-gap, and this emission grows exponentially and interferes with the PTC Cherenkov-like radiation. Naturally, this phenomena cannot be simply explained through Maxwell's equations. Next, we present the quantum model to complete the physical picture of light emission by a free-electron moving in a PTC.

To describe the underlying phenomena on a single-photon interaction level, we use canonical quantization tools. The resulting Hamiltonian of our system is (see details in SI [22]):

$$H_{tot} = H_{EM}(t) + H_e + H_I$$

$$= \sum_{k,\sigma} \hbar \frac{c|k|}{n_r} \left( \left( \frac{\epsilon_m(t)+1}{2\epsilon_m(t)} \right) \left( a^\dagger_{k\sigma} a_{k\sigma} + \frac{1}{2} \right) + \left( \frac{\epsilon_m(t)-1}{4\epsilon_m(t)} \right) \left( a_{k\sigma} a_{-k\sigma} + a^\dagger_{k\sigma} a^\dagger_{-k\sigma} \right) \right) + \frac{\widehat{P}^2}{2m} - \frac{e\widehat{P} \cdot \widehat{A}}{m} \quad (5)$$

with $n_r = \sqrt{\epsilon_r}$. This expression consists of the Hamiltonians for the EM field $H_{EM}$, the free electron energy $H_e$, and the interaction term $H_I$, where $n_r$ is the ambient refractive index, **k** and $\sigma$ are the wavenumber and polarization, $\epsilon(t)$ is the modulated permittivity, $a^\dagger_{k,\sigma}$, $a_{k,\sigma}$ are the creation and annihilation operators of a photon with $k$ and $\sigma$, $\widehat{\mathbf{P}}$ is the momentum operator of the electron and $\widehat{\mathbf{A}}$ is the vector potential operator of the EM field, written using $a_{k\sigma}$ (see SI [22]). Notice that $H_{EM}(t)$ contains terms describing the creation and annihilation of pairs of photons with opposite momenta $\hbar \mathbf{k}$ and $-\hbar \mathbf{k}$, highlighting the fact that the total momentum is conserved. We can therefore arrange the states in independent ladders of pairs of counter propagating photons with wavevectors **k** and −**k**, $|0_k, 0_{-k}\rangle$, $|1_k, 1_{-k}\rangle$, $|2_k, 2_{-k}\rangle$, etc. The last term of $H_{EM}$ couples these states by adding (or subtracting) precisely one pair of $k$ and $-k$ photons, thus not changing the overall momentum. For wavevectors residing in the gap, the number of photons increases exponentially, even if initially there were no such photons in the medium, i.e., these photons can appear spontaneously from vacuum fluctuations [23–25] – a pure QED effect.

To study the free-electron interaction with photons in such a medium, we add the terms of the electron energy $H_e$ and the interaction term $H_I$. As in the classical model, here too, we find the same novel type of radiation, in addition to the ordinary Cherenkov radiation. This radiation is caused by the exchange of energy quanta $\hbar\Omega$ between the electron and the PTC. The relation between the wavevector of the emitted photon and the electron velocity is simply given by energy and momentum conservation of the interaction process (see SI [22]), which results in $k_z\beta c = \omega - m\Omega$. This relation seems exactly equivalent to the classical relation we derived earlier; however, - from a QED perspective - $m$ is the number of energy quanta $\hbar\Omega$ delivered from the PTC to the emitted photon in the interaction process. Figure 3a shows the probability of the electron to emit a photon in a PTC in the subluminal regime, in a region of wavevectors located far from the momentum-gap. This radiation conforms at most wavevectors with the radiation we found using the classical analysis, based on Maxwell's equations (pink line). The emission probability is especially high when the energy-conservation condition for the $m = 1$ harmonic is fulfilled, i.e., $k_z\beta c = \omega - \Omega$. In this case, the electron contributes a quantum of energy $\hbar(\omega - \Omega)$ and emits a photon with energy $\hbar\omega$, which means that the PTC (the modulation of the refractive index) contributed energy of $\hbar\Omega$ to this interaction. The superluminal regime is even more interesting. When the velocity of the electron crosses the speed of light in the medium, an abrupt transition occurs in the radiation pattern: we find both the ordinary Cherenkov-type (shockwave) radiation at angle $\theta_{Ch} = \cos^{-1}(1/\beta n_{eff}(k))$ ($m = 0$) when the PTC does not contribute energy to the process, and an important additional outcome: the electron-PTC interaction with $m < 0$. In this latter case, the electron contributes the energy quanta $\hbar(\omega + |m|\Omega)$, but the energy of each emitted photon is still $\hbar\omega$. This means that when the electron emits radiation to these modes, it also returns energy back to the modulated medium. In this process, the electron slows down more than it does when it emits ordinary Cherenkov radiation in a stationary medium with the relation $\Delta\beta = \frac{(1-\beta_i^2)^{\frac{3}{4}}}{\beta_i m_0 c^2}\hbar(\omega + m\Omega)$, where $\beta_i$ is the initial velocity of the electron and $\Delta\beta$ is the change in the electron velocity due to the emission of a photon with energy $\hbar\omega$ and the loss of energy $\hbar m\Omega$ to the PTC (See SI for more details).

Another intriguing emission regime occurs in the momentum-gap region. In the momentum-gap, pairs of photons with opposite momentum are spontaneously emitted out of vacuum, similar to the dynamic Casimir effect where moving boundaries of a cavity create pairs of photons with half the modulation frequency [26]. Using Eq. 5 to study the quantum interaction of an electron and momentum-gap photons of the PTC, we find (Fig. 3b) that the photon emission rate increases exponentially (deep purple) compared to the emission rate without the PTC (pink). The exponential growth is shown by fitting the difference in emission rates to an exponential function (orange). This finding stands in sharp contrast to suppressed (prohibited) emission of photons in the bandgap of spatial photonic crystals [27]. We believe this is a general feature of emission into a time-varying medium: the emission rate is always higher for photons with wavenumbers $k$ inside a PTC momentum-gap.

In the same superluminal regime and under strong electron-photon coupling, characterized by the coupling strength $g$, we find an even more interesting effect at the angle opposite to the ordinary Cherenkov angle. In that direction, the final states of the emission process by the moving electron, $|P - \hbar k, n_k, n_{-k} - 1\rangle$, become quantum-degenerate with the states of the spontaneous emission created by vacuum fluctuations and enhanced by the PTC, $|P, n_k, n_{-k}\rangle$,. This quantum degeneracy of momentum and energy is lifted by the interaction term and causes avoided-crossing between the two processes, as shown by Fig. 4a. We mark the lines of electron radiation in a PTC with very weak electron-photon coupling (pink) and the bandgap wavevectors (orange). In principle, these lines would cross, but the strong interaction between the electron and the medium prevents the crossing. Figure 4b shows the growing emission probability at two new wavevectors, $k_{\text{gap}} \pm \Delta k$, where $\Delta k$ depends on the electron-photon interaction strength. Figure 4c compares the spontaneous two-photon emission over time, in the absence and in the presence of a free electron. When the electron is absent, the photon pairs are created in a rate that grows exponentially. The presence of the free-electron alters the process dramatically: the emission in the PTC momentum-gap is suppressed at the crossing point, highlighting the avoided crossing. Figure 4d, on the other hand, compares the radiation at the two wavevectors, with $k_\perp = k_{\text{gap}} \pm \Delta k$. It shows that emission at these wavevectors is enhanced due to the presence of the electron. These results are direct outcome of the underlying QED mechanism, as they cannot be explained through the Maxwell equations alone.

In conclusion, we analyzed the emission of radiation by a free-electron moving through a PTC, predicted the exponential enhancement of emission in the directions conforming to the momentum-gap, and pinpointed the outcome of QED in the form of avoided crossing. Our classical and quantum descriptions of the EM modes and electron interaction are general and can be applied to any variation of the permittivity in time, strong and abrupt as it may be. The recent progress on ultrafast epsilon-near-zero (ENZ) materials makes it promising to observe PTCs in the near future, and the utilization of ultrafast transmission electron microscopes (UTEM) [28] makes the ideas described here accessible to experiments [29]. Specifically, the recently developed ENZ materials with strong nonlinear effects were shown to possess very fast (fs) and strong (order of ~1) modulation of the permittivity in response to an ultrafast pulse [11,14,30]. In principle, a train of such pulses will result in time-periodic permittivity – creating a PTC., and then a free electron in this PTC will effectively acts as a relativistic EM emission source. This scheme can pave the way to new physics, such as a radiation scheme for relativistic dipoles known to show effects like the superlight inverse Doppler effect [31], or the study of quantum correlations between electrons and the photon pairs created by the modulation [32]. Looking forward, these ideas could also lead to novel tunable particle detectors, and to X-ray lasers drawing their energy from periodically-driven materials, which can be tuned over a large wavelength span.

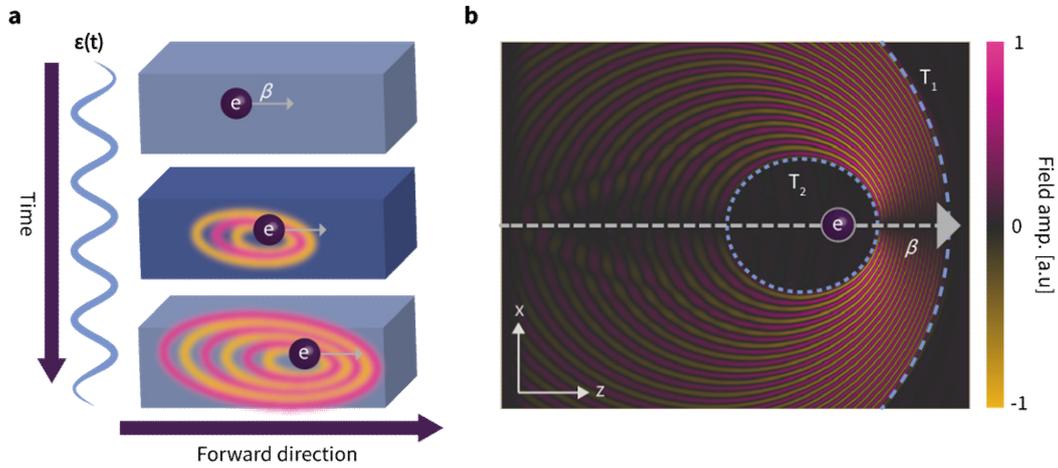

**Fig. 1. The process of free-electron radiation in a Photonic Time-Crystal. (a)** Schematics of a free electron moving and radiating in a PTC. When the index of refraction changes periodically in time, the electron moving in the medium emits EM radiation which depends on the modulation and on the electron velocity $\beta c$. **(b)** Magnetic field amplitude of the EM radiation, $H(x, y = 0, z, t)$, obtained by FDTD simulation for an electron moving at velocity $\beta$ in a PTC of 20 cycles that starts at time $t = T_1$ and ends at $t = T_2$; the snapshot is taken at $t > T_2$. High wavenumbers are emitted backwards. There is no radiation on the propagation axis because the polarization of the emitted waves cannot be perpendicular to the electron velocity.

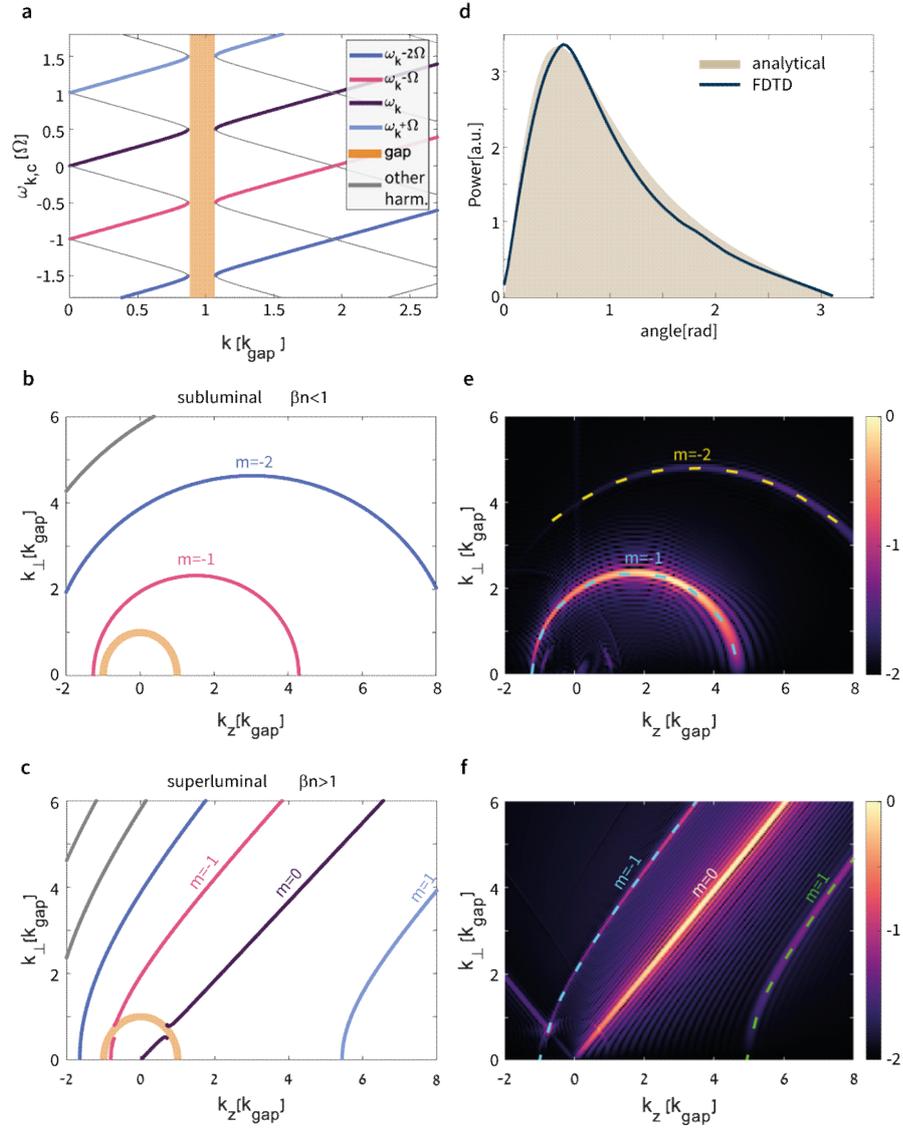

**Fig. 2. Band structure of the PTC and the radiation emitted by the free-electron.** **(a)** Unfolded band structure of the PTC, with the momentum gap marked in orange, and the various bands marked in different colors. Here, $k_{gap} = \Omega n_r / 2c_0$, where $\Omega$ is the PTC modulation frequency and $n_r = \sqrt{\epsilon_r}$ is the bias refractive index and the modulation strength is $\epsilon_1/\epsilon_0 = 0.4$. The changes of the permittivity in time cause reflections and refractions, that lead to the buildup of the Floquet band structure. **(b,c)** Phase-matched lines marking the directions of efficient radiation emitted by a free electron moving through a PTC, in the subluminal and in the superluminal regimes, respectively. **(d)** Power carried by the EM radiation vs. radiation angle: comparison between the analytic result and the FDTD simulation. **(e,f)** Radiation pattern calculated by FDTD simulations in the subluminal and superluminal regimes, respectively. The radiation lines correspond to the lines in (b,c).

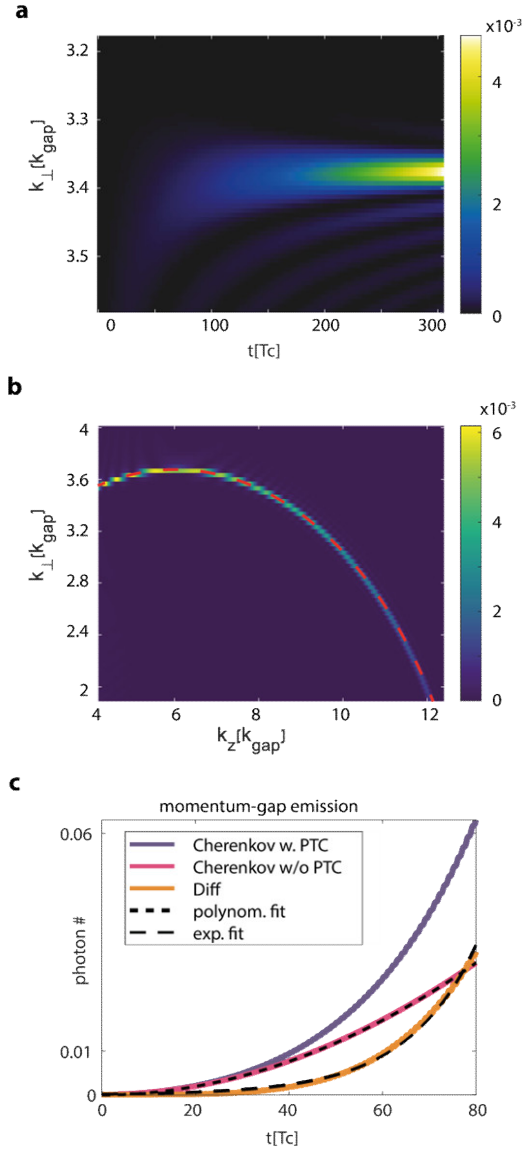

**Fig. 3. Free-electron interaction with single photons in a PTC. (a)** Photon emission probability as a function of time in a range of k-wavenumbers, emitted by a free electron moving in a PTC with $\beta = 0.6$, $\epsilon_1/\epsilon_0 = 0.1$ and $g/\omega = 7.8 \cdot 10^{-3}$. **(b)** Photon emission probability by a free electron ($\beta = 0.6$) moving in a PTC, as a function of the k-vector of the emitted photon, as observed after 300 PTC cycles. The dashed red line marks the angles of phase-matching between the electron and the EM waves (via the classical model), corresponding well to the calculated probability of emission. Both (a) and (b) are calculated with the Hamiltonian of Eq. 5. **(c)** Cherenkov radiation of an electron with $\beta = 0.9$ in a PTC (deep purple) compared to ordinary Cherenkov without a PTC (pink) as a function of time for wavenumbers residing in the momentum-gap center of the PTC, with $\epsilon(t) = 2 + 0.05\sin(\omega t)$ and coupling strength $g/\omega = 10^{-5}$. The difference between emission with and without the PTC (orange) follows an exponential fit (long-dashed black line). The number of photons emitted by the electron is exponentially enhanced, drawing energy from the temporal modulation of the refractive index.

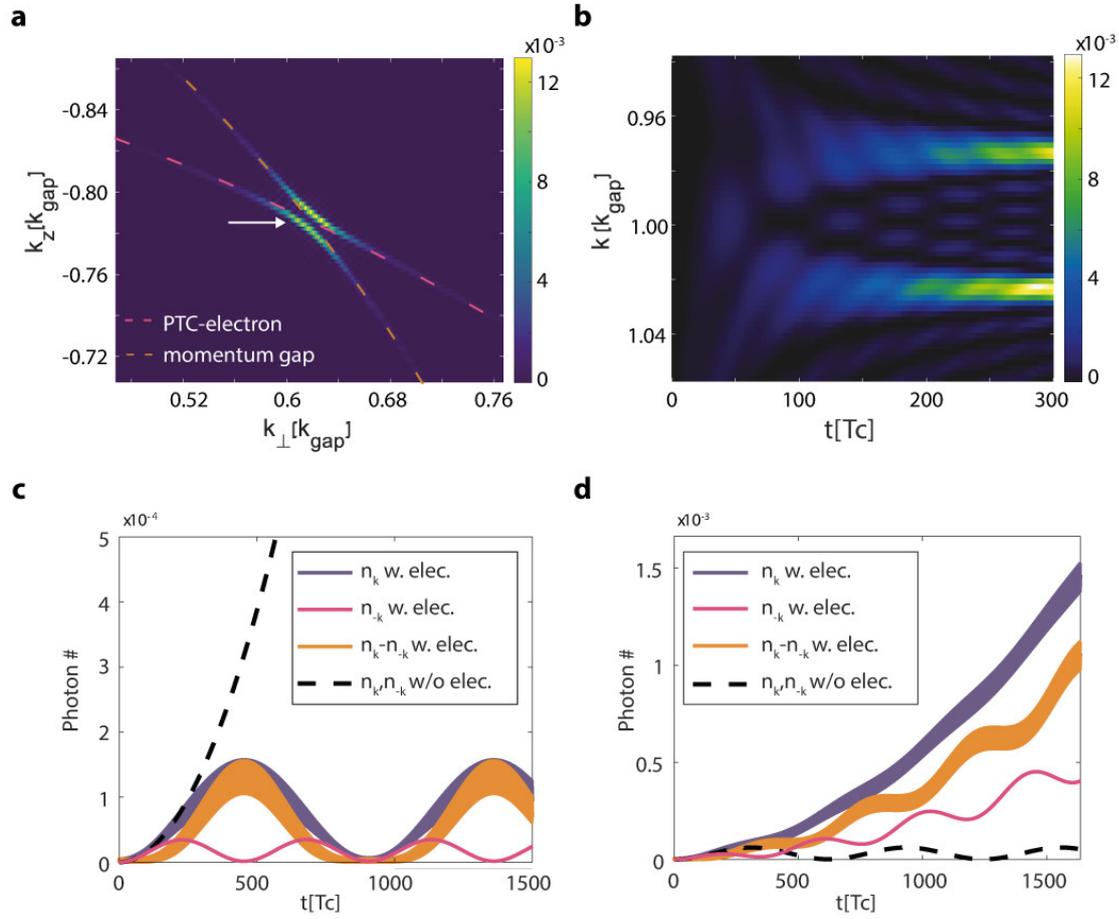

**Fig. 4. Free-electron back-emission near the momentum gap of a PTC ($\beta = 0.9$). (a)** Probability of the electron to emit a photon after 300 Tc. The EM radiation at the gap enhances the probability for the electron to emit a photon, which reaches a maximum at the momentum-gap (dashed pink), but this point coincides with the probability for spontaneously-emitted photon-pairs (dashed orange). The quantum interference between these two distinct effects results in avoided crossing. The size of the splitting depends on the light-matter interaction strength; here $g/\omega = 7.8 \cdot 10^{-3}$. **(b)** Probability of photon emission in the back-direction of the electron near the momentum-gap vs. time ($k$ wavevectors that belong to the dashed black line from (a)). The splitting becomes evident after several tens of modulation cycles in the permittivity. The electron-photon coupling strength is $g/\omega = 2.2 \cdot 10^{-2}$. **(c)** Comparison of the back-emission in the PTC momentum-gap with and without the electron. The presence of the electron suppresses the spontaneous two-photon emission in the gap. **(d)** Comparison of the back-emission in the band but near a PTC momentum-gap, at $k$-wavevector in the region of avoided crossing, with and without the electron: here the presence of the electron enhances the radiation emission. The coupling strength for (c),(d) is $g/\omega = 3 \cdot 10^{-3}$. We chose weak modulation amplitudes $\epsilon_1/\epsilon_0 = 2 \cdot 10^{-3}$ for (a),(b) and $10^{-4}$ for (c),(d) to stay in the perturbative limits of the quantum simulations.

# References

1. F. R. Morgenthaler, "Velocity Modulation of Electromagnetic Waves," IEEE Trans. Microw. Theory Tech. **6**, 167–172 (1958).

2. J. T. Mendonça and P. K. Shukla, "Time Refraction and Time Reflection: Two Basic Concepts," Phys. Scr. **65**, 160 (2002).

3. V. Bacot, M. Labousse, A. Eddi, M. Fink, and E. Fort, "Time reversal and holography with spacetime transformations," Nat. Phys. **12**, 972–977 (2016).

4. F. Biancalana, A. Amann, A. V. Uskov, and E. P. O'Reilly, "Dynamics of light propagation in spatiotemporal dielectric structures," Phys. Rev. E **75**, 046607 (2007).

5. A. B. Shvartsburg, "Optics of nonstationary media," Physics-Uspekhi **48**, 797 (2005).

6. D. E. Holberg and K. S. Kunz, "Parametric Properties of Fields in a Slab of Time-Varying Permittivity," IEEE Trans. Antennas Propag. **14**, 183–194 (1966).

7. J. R. Reyes-Ayona and P. Halevi, "Observation of genuine wave vector ($k$ or $\beta$) gap in a dynamic transmission line and temporal photonic crystals," Appl. Phys. Lett. **107**, 074101 (2015).

8. E. Lustig, Y. Sharabi, and M. Segev, "Topological aspects of photonic time crystals," Optica **5**, 1390 (2018).

9. Y. Sharabi, E. Lustig, and M. Segev, "Disordered Photonic Time Crystals," Phys. Rev. Lett. **126**, 163902 (2021).

10. A. Alù, M. G. Silveirinha, A. Salandrino, and N. Engheta, "Epsilon-near-zero metamaterials and electromagnetic sources: Tailoring the radiation phase pattern," Phys. Rev. B - Condens. Matter Mater. Phys. **75**, 155410 (2007).

11. M. Z. Alam, I. De Leon, and R. W. Boyd, "Large optical nonlinearity of indium tin oxide in its epsilon-near-zero region," Science **352**, 795–797 (2016).

12. L. Caspani, R. P. M. Kaipurath, M. Clerici, M. Ferrera, T. Roger, J. Kim, N. Kinsey, M. Pietrzyk, A. Di Falco, V. M. Shalaev, A. Boltasseva, and D. Faccio, "Enhanced Nonlinear Refractive Index in $\epsilon$-Near-Zero Materials," Phys. Rev. Lett. **116**, 233901 (2016).

13. Y. Zhou, M. Z. Alam, M. Karimi, J. Upham, O. Reshef, C. Liu, A. E. Willner, and R. W. Boyd, "Broadband frequency translation through time refraction in an epsilon-near-zero material," Nat. Commun. **11**, 1–7 (2020).

14. V. Bruno, C. Devault, S. Vezzoli, Z. Kudyshev, T. Huq, S. Mignuzzi, A. Jacassi, S. Saha, Y. D. Shah, S. A. Maier, D. R. S. Cumming, A. Boltasseva, M. Ferrera, M. Clerici, D. Faccio, R. Sapienza, and V. M. Shalaev, "Negative Refraction in Time-Varying Strongly Coupled Plasmonic-Antenna-Epsilon-Near-Zero Systems," Phys. Rev. Lett. **124**, 043902 (2020).

15. J. R. Zurita-Sánchez, P. Halevi, and J. C. Cervantes-González, "Reflection and transmission of a wave incident on a slab with a time-periodic dielectric function (t)," Phys. Rev. A - At. Mol. Opt. Phys. **79**, 053821 (2009).

16. J. R. Zurita-Sánchez, J. H. Abundis-Patiño, and P. Halevi, "Pulse propagation through a slab with time-periodic dielectric function ε(t)," Opt. Express **20**, 5586–5600 (2012).

17. P. A. Cherenkov, "Видимое свечение чистых жидкостей под действием γ-радиации," Uspekhi Fiz. Nauk (1967).

18. I. E. Tamm and I. Frank, "Coherent Visible Radiation of Fast Electrons Passing Through Matter," in *Selected Papers* (1991).

19. S. J. Smith and E. M. Purcell, "Visible light from localized surface charges moving across a grating," Phys. Rev. (1953).

20. We choose to work with the magnetic field because its mathematical link to the current source is free of time derivatives (Ampere's law).

21. S. M. Barnett and R. Loudon, "The enigma of optical momentum in a medium," Philos. Trans. R. Soc. A Math. Phys. Eng. Sci. **368**, 927–939 (2010).

22. *See Supplementary Materials.*

23. L. Parker, "Quantized Fields and Particle Creation in Expanding Universes. I," Phys. Rev. **183**, 1057–1068 (1969).

24. C. K. Law, "Effective Hamiltonian for the radiation in a cavity with a moving mirror and a time-varying dielectric medium," Phys. Rev. A **49**, 433–437 (1994).

25. T. Kawakubo and K. Yamamoto, "Photon creation in a resonant cavity with a nonstationary plasma mirror and its detection with Rydberg atoms," Phys. Rev. A **83**, 13819 (2011).

26. G. T. Moore, "Quantum Theory of the Electromagnetic Field in a Variable-Length One-Dimensional Cavity," J. Math. Phys. **11**, 2679–2691 (1970).

27. E. Yablonovitch, "Inhibited Spontaneous Emission in Solid-State Physics and Electronics," Phys. Rev. Lett. **58**, 2059–2062 (1987).

28. B. Barwick, D. J. Flannigan, and A. H. Zewail, "Photon-induced near-field electron microscopy," Nature **462**, 902–906 (2009).

29. R. Dahan, S. Nehemia, M. Shentcis, O. Reinhardt, Y. Adiv, X. Shi, O. Be'er, M. H. Lynch, Y. Kurman, K. Wang, and I. Kaminer, "Resonant phase-matching between a light wave and a free-electron wavefunction," Nat. Phys. **16**, 1123–1131 (2020).

30. A. Shaltout, A. Kildishev, and V. Shalaev, "Time-varying metasurfaces and Lorentz non-reciprocity," Opt. Mater. Express **5**, 2459–2467 (2015).

31. X. Shi, X. Lin, I. Kaminer, F. Gao, Z. Yang, J. D. Joannopoulos, M. Soljačić, and B. Zhang, "Superlight inverse Doppler effect," Nat. Phys. **14**, 1001–1005 (2018).

32. R. Dahan, A. Gorlach, U. Haeusler, A. Karnieli, O. Eyal, P. Yousefi, M. Segev, A. Arie, G. Eisenstein, P. Hommelhoff, and I. Kaminer, "Imprinting the quantum statistics of photons on free electrons," to appear (accepted), Science, (2021)